\documentclass[a4paper,11pt]{article}
\usepackage{pos}
\usepackage{dsfont}
\usepackage{slashed}
\usepackage{enumitem}

\renewcommand{\d}{\text{d}}

\title{Update on technical aspects of $B$ meson mixing at NNLO}

\author*[a]{Pascal Reeck}

\affiliation[a]{Institut für Theoretische Teilchenphysik, Karlsruhe Institute of Technology (KIT),\\
 Wolfgang-Gaede Straße 1, 76128 Karlsruhe, Germany}

\emailAdd{pascal.reeck@kit.edu}

\abstract{This report summarises important technical challenges and their solutions in the calculation of the next-to-next-to-leading order QCD corrections to the width difference $\Delta\Gamma_s$ in the $B_s-\overline{B}_s$ system. We focus on the treatment of spinor structures with products of up to 22 $\gamma$ matrices by means of projection and tensor integrals. Moreover, we present an approach for the semi-numeric computation of master integrals.}

\FullConference{Loops and Legs in Quantum Field Theory (LL2024)\\
 14-19, April, 2024\\
Wittenberg, Germany\\}

\begin{document}
\maketitle

\section{Introduction}
The mixing of $B_{s}$ and $\overline{B}_{s}$ mesons is fully described by the off-diagonal elements of the self-energy matrix $\Sigma$. Calculating these matrix elements leads to theoretical predictions for the mass difference $\Delta M_s$ and the width difference $\Delta \Gamma_s$ of the mass eigenstates. The width difference between $B_{s}$ and $\overline{B}_{s}$ mesons is given by
\begin{equation}
\begin{split}
\Delta \Gamma 
	&= - 2\lvert \Gamma_{12} \rvert \cos(\phi_{12}) + \mathcal{O}\left(\frac{\lvert\Gamma_{12}\rvert^2}{\lvert M_{12}\rvert^2}\right)\\
	&= \lvert \Sigma_{12} - \Sigma_{21}^* \rvert \cos(\phi_{12}) + \mathcal{O}\left(\frac{\lvert\Gamma_{12}\rvert^2}{\lvert M_{12}\rvert^2}\right),
\end{split}
\end{equation}
where the CP-violating phase $\phi_{12}$ is the phase difference between the phases of $M_{12}$ and $\Gamma_{12}$ (see, e.g.~Refs.~\cite{Gerlach:2024qlz, Lenz:2006hd} for more details). 

In our calculation, we perform the matching of a $\lvert\Delta B\rvert = 2$ matrix element calculated within effective $\lvert\Delta B\rvert = 1$ and $\lvert\Delta B\rvert = 2$ theories, where the high-energy and low-energy effects factorise into matching coefficients and operator matrix elements respectively. To obtain only the leading terms in $\Lambda_\text{QCD}/m_b$, the Heavy Quark Expansion (HQE) is used for the transition operator on the $\lvert\Delta B\rvert = 2$ side, which allows us to expand the operators in $\Lambda_\text{QCD}/m_b$ \cite{Khoze:1983yp, Shifman:1984wx, Khoze:1986fa, Chay:1990da, Bigi:1991ir, Bigi:1992su, Bigi:1993fe, Blok:1993va, Manohar:1993qn, Lenz:2014jha}. 

Matching the results of the $\lvert\Delta B\rvert = 1$ and $\lvert\Delta B\rvert = 2$ calculations leads to the matching coefficients $H^{ab}(z)$ and $\widetilde{H}^{ab}_S(z)$, which are needed to express the off-diagonal decay width element,
\begin{eqnarray}
	\Gamma_{12}^{ab} 
	&=& \frac{G_F^2m_b^2}{24\pi M_{B_s}} \left[ 
	H^{ab}(z)   \langle B_s|Q|\bar{B}_s \rangle
	+ \widetilde{H}^{ab}_S(z)  \langle B_s|\widetilde{Q}_S|\bar{B}_s \rangle
	\right]
	+ \mathcal{O}(\Lambda_{\rm QCD}/m_b).
	\label{eq::Gam^ab}
\end{eqnarray}
For more details and a complete definition of the operators, see Ref.~\cite{Reeck:2024iwk}. Compared to the experimental accuracies, theoretical predictions are still lagging behind, necessitating higher-order corrections to the previously known results, in particular the inclusion of penguin operators to NNLO. For a summary of current theoretical and experimental results, see Refs.~\cite{Gerlach:2022hoj, Dziurda:2024hrg}.

In this report, we will address a few of the technical challenges of this calculation: the treatment of products of up to 22 $\gamma$ matrices using either projectors or tensor reduction and the computation of master integrals.

\section{Projection of Dirac structures}

In this section, a practical implementation of an efficient algorithm for the projection of the Dirac spinor structures occurring in the calculation of mixing amplitudes is presented. For a more in-depth discussion and theoretical background see Ref.~\cite{Reeck:2024iwk}. 

This is done in two steps; first, the $\gamma$ matrices are ordered canonically and then the projectors are applied. Diagrams appearing in the calculation of mixing amplitudes from effective field theories contain spinor structures with a complicated tensor structure in Lorentz space since they are composed of a product of Dirac $\gamma$ matrices, which we will call \textit{spin lines} or \textit{Dirac chains}. Note that the chirality projector $P_{R/L}=(1\pm\gamma_5)/2$ can always be commuted to the end of a spin line since we are using the basis of Ref.~\cite{Chetyrkin:1997gb}. Thus, in our calculation, $\gamma_5$ never appears in closed fermion loops and we can use anti-commuting $\gamma_5$.

\begin{figure}
    \centering
    \includegraphics[scale=0.6]{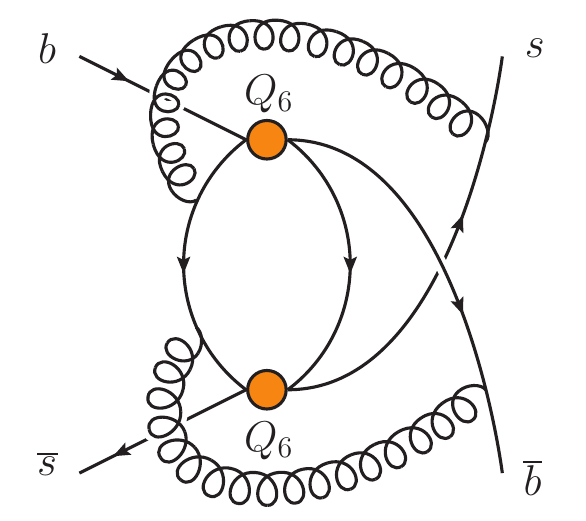}
    \caption{A diagram with one of the most complicated spinor structures that appear in the matching calculation. Note that the Lorentz indices corresponding to the effective operators $Q_6$ of the $\lvert\Delta B \rvert = 1$ theory are labelled as $\mu_i$ and $\nu_i$ while the gluons carry Lorentz indices $\alpha_i$.}
    \label{fig:nonplanar_3l}
\end{figure}

The maximum number of $\gamma$ matrices in our diagrams is eleven on each of the two spin lines, which includes at least one slashed momentum per spin line. We give an efficient algorithm for the treatment of such structures below, and for illustration purposes we will refer to the structures appearing at each step of the calculation of the diagram shown in Fig.~\ref{fig:nonplanar_3l}. We call $\gamma$ matrices which do not have their Lorentz index contracted with a propagator momentum \textit{pure}.
\begin{enumerate}[label=(\roman*)]
\item Project onto left-handed and right-handed spinor structures. It is sufficient to multiply with $P_{R/L} = (1\pm\gamma_5)/2$ and then drop all remaining terms containing $\gamma_5$. For the sample diagram, we have
\begin{equation}
    \gamma^{\mu_1} \gamma^{\mu_2} \gamma^{\mu_3} \slashed{k}_1 \gamma^{\alpha_1} \slashed{k}_2 \gamma^{\alpha_2} \slashed{k}_ 3\gamma^{\nu_1} \gamma^{\nu_2} \gamma^{\nu_3} \otimes \gamma_{\alpha_2} \slashed{k}_4 \gamma_{\mu_1} \gamma_{\mu_2} \gamma_{\mu_3} \slashed{k}_5 \gamma_{\nu_1} \gamma_{\nu_2} \gamma_{\nu_3} \slashed{k}_6 \gamma_{\alpha_1} + \text{3 terms}\,,
\end{equation}
where the $k_i$ are line momenta of the Feynman diagram in Fig.~\ref{fig:nonplanar_3l}.
\item Express all line momenta in terms of loop momenta and the external momentum, e.g.~for the sample diagram:
\begin{equation}
    \gamma^{\mu_1} \gamma^{\mu_2} \gamma^{\mu_3} \slashed{p}_3 \gamma^{\alpha_1} \slashed{p}_3 \gamma^{\alpha_2} \slashed{p}_2 \gamma^{\nu_1} \gamma^{\nu_2} \gamma^{\nu_3} \otimes \gamma_{\alpha_2} \slashed{p}_3 \gamma_{\mu_1} \gamma_{\mu_2} \gamma_{\mu_3} \slashed{p}_1 \gamma_{\nu_1} \gamma_{\nu_2} \gamma_{\nu_3} \slashed{p}_2 \gamma_{\alpha_1} + \text{95 terms}\,.
\end{equation}
\item Reorder the momenta on both spin lines to remove all duplicates and choose a unique ordering for each combination of momenta. The sample diagram then has the following spinor structure:
\begin{equation}
    \slashed{p}_3 \slashed{p}_2 \slashed{p}_1 \gamma^{\mu_2} \gamma^{\mu_3} \gamma^{\alpha_2} \gamma^{\nu_1} \gamma^{\nu_2} \gamma^{\nu_3} \otimes \slashed{p}_3 \gamma_{\alpha_2} \gamma_{\mu_2}\gamma_{\mu_3} \gamma_{\nu_1} \gamma_{\nu_2} \gamma_{\nu_3} + \text{181510 terms}
\end{equation}
\item Contract duplicate Lorentz indices on the same spin line after commuting the respective $\gamma$ matrices together.
\item Bring all pure $\gamma$ matrices into a canonical order. We use a lookup table which resolves the permutation of pure $\gamma$ matrices. For our example case, this leads to
\begin{equation}
    \slashed{p}_3 \slashed{p}_2 \slashed{p}_1 \gamma^{\rho_1} \gamma^{\rho_2} \gamma^{\rho_3} \gamma^{\rho_4} \gamma^{\rho_5} \gamma^{\rho_6} \otimes \slashed{p}_3 \gamma_{\rho_6} \gamma_{\rho_5}\gamma_{\rho_4} \gamma_{\rho_3} \gamma_{\rho_2} \gamma_{\rho_1}  + \text{40134 terms}.
\end{equation}
\item Map the terms with ordered pure $\gamma$ matrices and split-off slashed momenta directly onto the basis elements $B_i$ given in Ref.~\cite{Reeck:2024iwk}. This is done by using another lookup table, which has been calculated using projectors on the ordered gamma structures. When calculating the lookup table, we can also choose to use different projectors, i.e.~different scalar products, depending on the spinor structure which needs to be resolved. For details on the chosen scalar products see Ref.~\cite{Reeck:2024iwk}. In the end, our sample diagram results in 
\begin{equation}
    \frac{(p_1\cdot p_3)^2 p_2^2}{d^3 - 6d ^2 + 11 d - 6} \times B_{45}  + \text{104335 terms},
\end{equation}
where $d=4-2\epsilon$ is the spacetime dimension, and the basis element $B_{45}$ is defined as 
\begin{equation}
B_{45} =  \gamma_{\mu_1} \dots \gamma_{\mu_{11}}  \otimes  \gamma^{\mu_{11}} \dots \gamma^{\mu_1}.
\end{equation}
\item Map basis elements onto operator matrix elements, e.g.~$\langle B_s|Q|\bar{B}_s \rangle$, $\langle B_s|\widetilde{Q}_S|\bar{B}_s \rangle$, $\langle B_s| E^{(1)}_1|\bar{B}_s \rangle$, etc. This can be done using a lookup table.
\end{enumerate}

\section{Tensor integrals}

Tensor reduction provides an alternative method for evaluating amplitudes with a non-trivial Lorentz structure. Unlike the projector technique, it does not rely on assumptions about the Dirac and colour structures that might emerge in the resulting expressions. Additionally, this approach does not suffer from performance issues resulting from traces over a large number of $\gamma$ matrices.

Nevertheless, tensor reduction formulae become increasingly complex as the tensor rank and the number of loop momenta grow. Our calculations require handling three-loop tensors of up to rank eleven with a single external momentum.

A straightforward way of deriving tensor reduction formulae consists of writing down the most general ansatz containing all tensor structures allowed by the symmetries. For example, 
in the case of a rank-three two-loop integral with one external momentum $q$, the tensor integral can be decomposed as
\begin{equation}
\begin{split}
\int  p_1^\mu p_2^\nu p_2^\rho &f(p_1,p_2,q)\, \d^d p_1 \, \d^d p_2 =\\
& \int \left ( g^{\mu \nu} q^\rho c_1 + g^{\mu \rho} q^\nu c_2  + g^{\nu \rho} q^\mu c_3 + q^\mu q^\nu q^\rho c_4 \right )  f(p_1,p_2,q) \, \d^d p_1 \, \d^d p_2.
\label{eq:tdec}
\end{split}
\end{equation}
By contracting Eq.~\eqref{eq:tdec} with each of the tensor structures multiplying $c_i$, one obtains a linear system of equations which can be solved for the coefficients $c_i$. 

More complex tensor reductions require dedicated software to make them tractable and we would like to highlight one particularly attractive solution. An efficient algorithm which makes use of some of the underlying symmetries is presented in Ref.~\cite{Pak:2011xt}. This algorithm has been implemented in the routine \texttt{Tdec} available in \textsc{FeynCalc} 10 \cite{Mertig:1990an,Shtabovenko:2016sxi,Shtabovenko:2020gxv,Shtabovenko:2023idz}, which can also interface to the fast linear solver \textsc{Fermat} \cite{Lewis:Fermat} using the development version of the \textsc{FeynHelpers}\footnote{\url{https://github.com/FeynCalc/feynhelpers}} add-on \cite{Shtabovenko:2016whf}.

These tools provide a powerful check of the results obtained using projectors for a large subclass of diagrams. The next step in the calculation is to reduce the scalar integrals to master integrals. The computation of these is discussed in the next section.

\section{Calculation of master integrals}

To cover the entire physical range of $z=m_c^2/m_b^2$, the master integrals need to be extended beyond the naive Taylor expansion of the $|\Delta B| = 1$ amplitudes up to $\mathcal{O}(z)$ that was used in the results published in Refs.~\cite{Gerlach:2022wgb,Gerlach:2022hoj}. Here, we describe a semi-numerical method we use to obtain results to very high orders in $z$ for all relevant integrals at two and three loops. Allowing for different renormalisation schemes of the quark masses, we need to cover a range of $x\equiv\sqrt{z} \in [0.20, 0.43]$.

The method used to obtain semi-numerical results is the so-called ``expand and match'' approach~\cite{Fael:2021kyg,Fael:2022rgm,Fael:2022miw,Fael:2023zqr}. This method uses the differential matrix equation in $x$ of the master integrals to produce a series expansion in the kinematic variable with numerical coefficients. To cover the entire range of relevant mass ratios, expansions are computed around different points $x_E$ which are then matched to boundary conditions of these integrals at points $x_M$ sufficiently close to $x_E$. If there is no threshold at $x_E$, the master integrals are regular at this point and it is even possible to choose $x_M=x_E$. In any case, on can use either analytical or numerical boundary conditions, but we opt for the numerical approach.

The most general ansatz that we need for the master integrals appearing in our diagrams can be written as
\begin{eqnarray}
  M_i(x, \epsilon) &=&
  \sum_{j=-2}^{\epsilon_{\mathrm{max}}}
  \sum_{m}^{j+2}  \sum_{n}^{n_{\mathrm{max}}}
  c_{i,j,m,n} \epsilon^j \, (x-x_E)^{n/2} \,
                          \log^m\left(x-x_E\right)\,,
  \label{eq::ansatz}
\end{eqnarray}
where we have used that in the imaginary part of three-loop two-point functions one has at most $1/\epsilon^2$ poles. For details on the limits of the summation variables see Ref.~\cite{Reeck:2024iwk}. Note that depending on whether the expansion point is regular or singular, the ansatz can be simplified. For the boundary conditions we generate numerical results using {\tt AMFlow}~\cite{Liu:2022chg} with a precision of 100 digits. 

With these results, we obtain a precision of the integral over the entire physical range of 20 digits or more. To check the validity of our results, we compare the coefficients of the $z^0$ and $z^1$ terms of the matching coefficients $H^{ab}$ and $\widetilde{H}^{ab}_S$ introduced in Eq.~(\ref{eq::Gam^ab}) to the analytic results from Ref.~\cite{Gerlach:2022hoj}. The agreement is better than $10^{-40}$.

As an illustration, we give the non-fermionic (i.e. $n_f=0$) NNLO contribution of two $Q_1$ insertions to the matching coefficient $H^{cc}(x)$. The expansion for $x\to0$ up to $x^{10}$ reads:
\begin{eqnarray}
   H^{cc, (2)}(x) &=& \Big(61.49166 + (334.9196 - 1227.224 \log x + 88.00000 \log^{2} x) \,x^2 - 11.69731 \,x^3
    \nonumber\\&&\mbox{}
    + (-4522.778 + 1030.276 \log x + 376.7253 \log^{2} x) \,x^4 + 199.3985 \,x^5 \nonumber\\&&\mbox{}
    + (-5312.014 + 6117.164 \log x + 1408.251 \log^{2} x) \,x^6 + 500.4536 \,x^7 \nonumber\\&&\mbox{}
    + (-1386.574 + 13388.09 \log x - 773.9628 \log^{2} x) \,x^8 + 1591.938 \,x^9 \nonumber\\&&\mbox{}
    + (-82.72167 + 27995.10 \log x - 1593.620 \log^{2} x) \,x^{10} +{\cal O}(x^{11})\Big)\, C_{1}^2\,,
\end{eqnarray}
where the renormalisation scale is set to $\mu=m_b$, and the quark masses are renormalised in the pole scheme. In defining the $\Delta B = 2$ evanescent operators, we have followed Ref.~\cite{Asatrian:2017qaz} and have set all other $\mathcal{O}(\epsilon)$ contributions of physical operators to evanescent operators that were not previously defined to zero.

\section{Conclusion}

Major technical challenges a higher-order calculation of $B_s-\overline{B}_s$ mixing have been addressed and their solutions presented. A reduction of Dirac structures with up to 22 $\gamma$ matrices on either of the two spin lines is now tractable, and the computation of semi-analytic three-loop master integrals with a dependence on the ratio of the charm and bottom quark masses has been carried out for most of the relevant diagrams. Moreover, the methods presented here can also be applied to other processes involving $B$ and $D$ mesons, e.g.~for their decay rates.

\section*{Acknowledgments}

I would like to thank Ulrich Nierste, Vladyslav Shtabovenko and Matthias Steinhauser for their support and collaboration on the topics presented here. This research was supported by the Deutsche Forschungsgemeinschaft (DFG, German Research Foundation) under grant 396021762 — TRR 257 ``Particle Physics Phenomenology after the Higgs Discovery''.

\bibliographystyle{JHEP}
\bibliography{references}

\providecommand{\href}[2]{#2}\begingroup\raggedright\begin{thebibliography}{10}

\bibitem{Gerlach:2024qlz}
M.~Gerlach, U.~Nierste, P.~Reeck, V.~Shtabovenko and M.~Steinhauser,
  \emph{{NNLO QCD corrections to $\Delta \Gamma_s$ in the $B_s-\overline{B}_s$
  system}},  in \emph{{12th International Workshop on the CKM Unitarity
  Triangle}}, 3, 2024 [\href{https://arxiv.org/abs/2403.08316}{{\ttfamily
  2403.08316}}].

\bibitem{Lenz:2006hd}
A.~Lenz and U.~Nierste, \emph{{Theoretical update of $B_s - \bar{B}_s$
  mixing}}, \href{https://doi.org/10.1088/1126-6708/2007/06/072}{\emph{JHEP}
  {\bfseries 06} (2007) 072}
  [\href{https://arxiv.org/abs/hep-ph/0612167}{{\ttfamily hep-ph/0612167}}].

\bibitem{Khoze:1983yp}
V.A.~Khoze and M.A.~Shifman, \emph{{HEAVY QUARKS}},
  \href{https://doi.org/10.1070/PU1983v026n05ABEH004398}{\emph{Sov. Phys. Usp.}
  {\bfseries 26} (1983) 387}.

\bibitem{Shifman:1984wx}
M.A.~Shifman and M.B.~Voloshin, \emph{{Preasymptotic Effects in Inclusive Weak
  Decays of Charmed Particles}}, {\emph{Sov. J. Nucl. Phys.} {\bfseries 41}
  (1985) 120}.

\bibitem{Khoze:1986fa}
V.A.~Khoze, M.A.~Shifman, N.G.~Uraltsev and M.B.~Voloshin, \emph{{On Inclusive
  Hadronic Widths of Beautiful Particles}}, {\emph{Sov. J. Nucl. Phys.}
  {\bfseries 46} (1987) 112}.

\bibitem{Chay:1990da}
J.~Chay, H.~Georgi and B.~Grinstein, \emph{{Lepton energy distributions in
  heavy meson decays from QCD}},
  \href{https://doi.org/10.1016/0370-2693(90)90916-T}{\emph{Phys. Lett. B}
  {\bfseries 247} (1990) 399}.

\bibitem{Bigi:1991ir}
I.I.Y.~Bigi and N.G.~Uraltsev, \emph{{Gluonic enhancements in non-spectator
  beauty decays: An Inclusive mirage though an exclusive possibility}},
  \href{https://doi.org/10.1016/0370-2693(92)90066-D}{\emph{Phys. Lett. B}
  {\bfseries 280} (1992) 271}.

\bibitem{Bigi:1992su}
I.I.Y.~Bigi, N.G.~Uraltsev and A.I.~Vainshtein, \emph{{Nonperturbative
  corrections to inclusive beauty and charm decays: QCD versus phenomenological
  models}}, \href{https://doi.org/10.1016/0370-2693(92)90908-M}{\emph{Phys.
  Lett. B} {\bfseries 293} (1992) 430}
  [\href{https://arxiv.org/abs/hep-ph/9207214}{{\ttfamily hep-ph/9207214}}].

\bibitem{Bigi:1993fe}
I.I.Y.~Bigi, M.A.~Shifman, N.G.~Uraltsev and A.I.~Vainshtein, \emph{{QCD
  predictions for lepton spectra in inclusive heavy flavor decays}},
  \href{https://doi.org/10.1103/PhysRevLett.71.496}{\emph{Phys. Rev. Lett.}
  {\bfseries 71} (1993) 496}
  [\href{https://arxiv.org/abs/hep-ph/9304225}{{\ttfamily hep-ph/9304225}}].

\bibitem{Blok:1993va}
B.~Blok, L.~Koyrakh, M.A.~Shifman and A.I.~Vainshtein, \emph{{Differential
  distributions in semileptonic decays of the heavy flavors in QCD}},
  \href{https://doi.org/10.1103/PhysRevD.50.3572}{\emph{Phys. Rev. D}
  {\bfseries 49} (1994) 3356}
  [\href{https://arxiv.org/abs/hep-ph/9307247}{{\ttfamily hep-ph/9307247}}].

\bibitem{Manohar:1993qn}
A.V.~Manohar and M.B.~Wise, \emph{{Inclusive semileptonic B and polarized
  Lambda(b) decays from QCD}},
  \href{https://doi.org/10.1103/PhysRevD.49.1310}{\emph{Phys. Rev. D}
  {\bfseries 49} (1994) 1310}
  [\href{https://arxiv.org/abs/hep-ph/9308246}{{\ttfamily hep-ph/9308246}}].

\bibitem{Lenz:2014jha}
A.~Lenz, \emph{{Lifetimes and heavy quark expansion}},
  \href{https://doi.org/10.1142/S0217751X15430058}{\emph{Int. J. Mod. Phys. A}
  {\bfseries 30} (2015) 1543005}
  [\href{https://arxiv.org/abs/1405.3601}{{\ttfamily 1405.3601}}].

\bibitem{Reeck:2024iwk}
P.~Reeck, V.~Shtabovenko and M.~Steinhauser, \emph{{$B$ meson mixing at NNLO:
  technical aspects}},  \href{https://arxiv.org/abs/2405.14698}{{\ttfamily
  2405.14698}}.

\bibitem{Gerlach:2022hoj}
M.~Gerlach, U.~Nierste, V.~Shtabovenko and M.~Steinhauser, \emph{{Width
  Difference in the B-B\textasciimacron{} System at Next-to-Next-to-Leading
  Order of QCD}},
  \href{https://doi.org/10.1103/PhysRevLett.129.102001}{\emph{Phys. Rev. Lett.}
  {\bfseries 129} (2022) 102001}
  [\href{https://arxiv.org/abs/2205.07907}{{\ttfamily 2205.07907}}].

\bibitem{Dziurda:2024hrg}
A.~Dziurda et~al., \emph{{Summary of Working Group 4: Mixing and mixing-related
  $CP$ violation in the B system: $\Delta m$, $\Delta \Gamma$, $\phi_s$,
  $\phi_1/\alpha$, $\phi_2/\beta$, $\phi_3/\gamma$}},  in \emph{{12th
  International Workshop on the CKM Unitarity Triangle}}, 4, 2024
  [\href{https://arxiv.org/abs/2404.03945}{{\ttfamily 2404.03945}}].

\bibitem{Chetyrkin:1997gb}
K.G.~Chetyrkin, M.~Misiak and M.~Munz, \emph{{$|\Delta F| = 1$ nonleptonic
  effective Hamiltonian in a simpler scheme}},
  \href{https://doi.org/10.1016/S0550-3213(98)00131-X}{\emph{Nucl. Phys. B}
  {\bfseries 520} (1998) 279}
  [\href{https://arxiv.org/abs/hep-ph/9711280}{{\ttfamily hep-ph/9711280}}].

\bibitem{Pak:2011xt}
A.~Pak, \emph{{The Toolbox of modern multi-loop calculations: novel analytic
  and semi-analytic techniques}},
  \href{https://doi.org/10.1088/1742-6596/368/1/012049}{\emph{J. Phys. Conf.
  Ser.} {\bfseries 368} (2012) 012049}
  [\href{https://arxiv.org/abs/1111.0868}{{\ttfamily 1111.0868}}].

\bibitem{Mertig:1990an}
R.~Mertig, M.~Bohm and A.~Denner, \emph{{FEYN CALC: Computer algebraic
  calculation of Feynman amplitudes}},
  \href{https://doi.org/10.1016/0010-4655(91)90130-D}{\emph{Comput. Phys.
  Commun.} {\bfseries 64} (1991) 345}.

\bibitem{Shtabovenko:2016sxi}
V.~Shtabovenko, R.~Mertig and F.~Orellana, \emph{{New Developments in FeynCalc
  9.0}}, \href{https://doi.org/10.1016/j.cpc.2016.06.008}{\emph{Comput. Phys.
  Commun.} {\bfseries 207} (2016) 432}
  [\href{https://arxiv.org/abs/1601.01167}{{\ttfamily 1601.01167}}].

\bibitem{Shtabovenko:2020gxv}
V.~Shtabovenko, R.~Mertig and F.~Orellana, \emph{{FeynCalc 9.3: New features
  and improvements}},
  \href{https://doi.org/10.1016/j.cpc.2020.107478}{\emph{Comput. Phys. Commun.}
  {\bfseries 256} (2020) 107478}
  [\href{https://arxiv.org/abs/2001.04407}{{\ttfamily 2001.04407}}].

\bibitem{Shtabovenko:2023idz}
V.~Shtabovenko, R.~Mertig and F.~Orellana, \emph{{FeynCalc 10: Do multiloop
  integrals dream of computer codes?}},
  \href{https://arxiv.org/abs/2312.14089}{{\ttfamily 2312.14089}}.

\bibitem{Lewis:Fermat}
R.~Lewis, \emph{{FERMAT}}, {\emph{\url{https://home.bway.net/lewis}} }.

\bibitem{Shtabovenko:2016whf}
V.~Shtabovenko, \emph{{FeynHelpers: Connecting FeynCalc to FIRE and
  Package-X}}, \href{https://doi.org/10.1016/j.cpc.2017.04.014}{\emph{Comput.
  Phys. Commun.} {\bfseries 218} (2017) 48}
  [\href{https://arxiv.org/abs/1611.06793}{{\ttfamily 1611.06793}}].

\bibitem{Gerlach:2022wgb}
M.~Gerlach, U.~Nierste, V.~Shtabovenko and M.~Steinhauser, \emph{{The width
  difference in $B - \bar B$ mixing at order $\alpha_s$ and beyond}},
  \href{https://doi.org/10.1007/JHEP04(2022)006}{\emph{JHEP} {\bfseries 04}
  (2022) 006} [\href{https://arxiv.org/abs/2202.12305}{{\ttfamily
  2202.12305}}].

\bibitem{Fael:2021kyg}
M.~Fael, F.~Lange, K.~Sch\"onwald and M.~Steinhauser, \emph{{A semi-analytic
  method to compute Feynman integrals applied to four-loop corrections to the $
  \overline{\mathrm{MS}} $-pole quark mass relation}},
  \href{https://doi.org/10.1007/JHEP09(2021)152}{\emph{JHEP} {\bfseries 09}
  (2021) 152} [\href{https://arxiv.org/abs/2106.05296}{{\ttfamily
  2106.05296}}].

\bibitem{Fael:2022rgm}
M.~Fael, F.~Lange, K.~Sch\"onwald and M.~Steinhauser, \emph{{Massive Vector
  Form Factors to Three Loops}},
  \href{https://doi.org/10.1103/PhysRevLett.128.172003}{\emph{Phys. Rev. Lett.}
  {\bfseries 128} (2022) 172003}
  [\href{https://arxiv.org/abs/2202.05276}{{\ttfamily 2202.05276}}].

\bibitem{Fael:2022miw}
M.~Fael, F.~Lange, K.~Sch\"onwald and M.~Steinhauser, \emph{{Singlet and
  nonsinglet three-loop massive form factors}},
  \href{https://doi.org/10.1103/PhysRevD.106.034029}{\emph{Phys. Rev. D}
  {\bfseries 106} (2022) 034029}
  [\href{https://arxiv.org/abs/2207.00027}{{\ttfamily 2207.00027}}].

\bibitem{Fael:2023zqr}
M.~Fael, F.~Lange, K.~Sch\"onwald and M.~Steinhauser, \emph{{Massive three-loop
  form factors: Anomaly contribution}},
  \href{https://doi.org/10.1103/PhysRevD.107.094017}{\emph{Phys. Rev. D}
  {\bfseries 107} (2023) 094017}
  [\href{https://arxiv.org/abs/2302.00693}{{\ttfamily 2302.00693}}].

\bibitem{Liu:2022chg}
X.~Liu and Y.-Q.~Ma, \emph{{AMFlow: A Mathematica package for Feynman integrals
  computation via auxiliary mass flow}},
  \href{https://doi.org/10.1016/j.cpc.2022.108565}{\emph{Comput. Phys. Commun.}
  {\bfseries 283} (2023) 108565}
  [\href{https://arxiv.org/abs/2201.11669}{{\ttfamily 2201.11669}}].

\bibitem{Asatrian:2017qaz}
H.M.~Asatrian, A.~Hovhannisyan, U.~Nierste and A.~Yeghiazaryan, \emph{{Towards
  next-to-next-to-leading-log accuracy for the width difference in the
  $B_s-\bar{B}_s$ system: fermionic contributions to order $(m_c/m_b)^0$ and
  $(m_c/m_b)^1$}}, \href{https://doi.org/10.1007/JHEP10(2017)191}{\emph{JHEP}
  {\bfseries 10} (2017) 191}
  [\href{https://arxiv.org/abs/1709.02160}{{\ttfamily 1709.02160}}].

\end{thebibliography}\endgroup

\end{document}